\documentclass[prd,aps,twocolumn,preprintnumbers,amsmath,amssymb,nofootinbib,superscriptaddress,notitlepage]{revtex4-1}

\usepackage{epsfig}
\usepackage[utf8]{inputenc}
\usepackage{color}
\usepackage[colorlinks=true,
linkcolor=blue,
breaklinks=true,
urlcolor=blue,
citecolor=blue]{hyperref}

\usepackage{graphicx}

\newcommand{\nn}{\nonumber}
\newcommand{\be}{\begin{equation}}
\newcommand{\ee}{\end{equation}}
\newcommand{\bea}{\begin{eqnarray}}
\newcommand{\eea}{\end{eqnarray}}
\newcommand{\ba}{\begin{array}}
\newcommand{\ea}{\end{array}}
\newcommand{\bi}{\begin{itemize}}
\newcommand{\ei}{\end{itemize}}

\newcommand{\lf}{\left}
\newcommand{\rg}{\right}

\newcommand{\ucas}{\affiliation{University of Chinese Academy of Sciences, Beijing 100049, China}}

\newcommand{\imp}{\affiliation{Institute of Modern Physics, Chinese Academy of Sciences, Lanzhou 730000, China}}

\newcommand{\physUCAS}{\affiliation{School of Physical Sciences, University of Chinese Academy of Sciences (UCAS), Beijing 100049, China}}

\newcommand{\liaoning}{\affiliation{School of Physics, Liaoning University, Shenyang 110036, China}}

\newcommand{\sucas}{\affiliation{Southern Center for Nuclear-Science Theory (SCNT), Institute of Modern Physics,
Chinese Academy of Sciences, Huizhou 516000, Guangdong Province, China}}

\begin{document}

\title{Exclusive Charmonium Production at the Electron-ion collider in China}

\author{Xue Wang}\email{wangxue@impcas.ac.cn}
\liaoning
\imp

\author{Xu Cao}\email{caoxu@impcas.ac.cn}
\imp
\ucas

\author{Aiqiang Guo}\email{guoaq@impcas.ac.cn}
\imp
\ucas


\author{Li Gong}\email{gongli@lnu.edu.cn}
\liaoning

\author{Xiao-Shen Kang}\email{kangxiaoshen@lnu.edu.cn}
\liaoning

\author{Yu-Tie Liang}\email{liangyt@impcas.ac.cn}
\imp
\ucas


\author{Jia-Jun Wu}\email{wujiajun@ucas.ac.cn}
\physUCAS\sucas

\author{Ya-Ping Xie}\email{xieyaping@impcas.ac.cn}
\imp
\ucas

\date{\today}

\begin{abstract}
    %
We investigate the exclusive $J/\psi$ production at the future Electron-ion collider in China by utilizing the eSTARlight event generator.
We model the cross-section and kinematics by fitting to the world data of $J/\psi$ photoproduction.
Projected statistical uncertainties on $J/\psi$ production are based on the design of a central detector, which consists of a tracker and vertex subsystem.
The precision of the pseudo-data allows us to probe the near-threshold mechanism, e.g. the re-scattering effect.
The significance of the forward amplitudes is discussed as well.
The design and optimization of the detector enhance the potential for exploring the near-threshold region and the realm of high four-momentum transfer squared, which is of particular interest on several physics topics.

\end{abstract}

\maketitle

\section{Introduction} \label{sec:intro}

Exclusive photo- and electro-production of heavy quarkonium in different kinematic regions are expected to probe a rich of physical topics with extensive coverage of quantum chromodynamics (QCD) dynamics at short distances and QCD-inspired models \cite{Cao:2023rhu}.
At high energies the heavy quarkonium mesons produced through soft and hard pomeron exchanges are sensitive to the gluon distribution function \cite{Sibirtsev:2004ca,Jones:2016ldq}.
They provide an effective way to obtain the transverse spatial distribution of sea quarks and gluons,
one of the main goals at the Electron Ion Collider (EIC) in Brookhaven National Laboratory \cite{AbdulKhalek:2021gbh}.

On the other hand, what mechanism is responsible for heavy quarkonium photoproduction at low energies remains a matter of controversy~\cite{Lee:2022ymp}.
The radiative decay of heavy quarkonium to light mesons is very small, unlike the light vector mesons production by real and virtual photons.
Thus the single light meson exchange contribution to heavy quarkonium photoproduction is expected to be insignificant.
%
%
In Ref.~\cite{Sibirtsev:2004ca},
they found that from a systematic analysis of the available experimental data, the mechanism at low energies and large-$t$ region might differ from Pomeron or two-gluon exchange.
Thus, the soft Pomeron exchange may not dominate starting from the reaction threshold.
%
%
The three-gluon exchange was proposed to dominate near threshold \cite{Brodsky:2000zc} and those of symmetric color configuration is recently argued to vanish because of the $C$-parity conservation \cite{Sun:2021pyw,Sun:2021gmi}.

Relying on the Vector-Meson Dominance (VMD) assumption, the near-threshold domain gives access to the quarkonium-nucleon interaction characterized by scattering length.
The spin-averaged $s$-wave scattering length is at the level of several tens milli femto meter (mfm) extracted in a dispersive analysis \cite{Gryniuk:2016mpk} and several mfm evaluated by a momentum expansion \cite{Strakovsky:2021vyk,Cao:2023rhu}.
%
Therefore, nucleon is nearly transparent, permitting the uninterrupted passage of heavy quarkonium.
%
%
The validity of extending the VMD model to encompass heavy vector mesons remains a contentious issue.
Since the heavy vector mesons coupling directly to the photo is deeply off-shell within VMD model  \cite{Wu:2019adv}, how to correctly include the off-shell effect is essential in the estimation of the strength and momentum dependence of the transition from photons to heavy vector mesons \cite{Xu:2021mju}.
%
An alternative production mechanism has been suggested, which involves the rescattering process through open charm intermediate states and is distinctively characterized by cusp effects observed in production cross sections \cite{Du:2020bqj}.
The situation becomes more complex with the potential emergence of pentaquark states, whose masses are located in the region close to the threshold \cite{Wang:2015jsa,HillerBlin:2016odx,Cao:2019kst,Wu:2019adv}.
These anticipated exotic states are expected to couple with considerable strength to $J/\psi p$ \cite{Wu:2010jy,Huang:2018wed}, a crucial channel for understanding the internal composition of these exotic states \cite{Guo:2017jvc}.

Another interesting physics aspect is taking advantage of the photo- and electro-production of heavy quarkonium as a means of revealing the origin of the proton mass \cite{Kharzeev:1995ij,Kharzeev:1998bz}.
The early intention via the QCD trace anomaly based on the QCD multipole expansion near the threshold is critically reviewed by explicit perturbative QCD calculations \cite{Sun:2021pyw,Sun:2021gmi}.
It seems that the large-$t$ region is connected to the gluonic gravitational form
factors (GFFs) of the nucleon under some approximations in a holographic QCD analysis \cite{Hatta:2018ina,Mamo:2019mka} and in a Generalized Parton Distribution (GPD) formalism \cite{Guo:2021ibg,Guo:2023pqw,Guo:2023qgu}.
%
%
The domain of large photon virtualities and small-$t$ are carefully examined as well \cite{Boussarie:2020vmu,Hatta:2018ina}. 
The GFFs can be used to decipher the mechanical properties of nucleon,
e.g. mechanical radius, pressure and shear force distributions.

\begin{figure}
  \centering
    \includegraphics[scale=0.90]{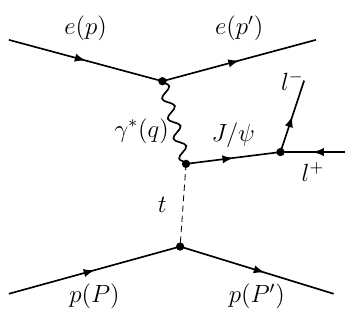}
  \caption{Schematic $t-$channel diagram of the electroproduction process $ep \rightarrow e^-p J/\psi \rightarrow e^-p \mu^{+}\mu^{-}$ (or $ep e^{+}e^{-}$)
  The relevant kinematic variables are labeled besides the lines with virtuality of photon $Q^2 = -q^2$ and four momentum transfer squared $t = (P^\prime - P)^2$.
  The invariant mass of the photon-proton system is defined as $W = \sqrt{(q+P)^2}$ and the total c.m energy as $s=(p + P)^2$.}
  \label{fig:feynman}
\end{figure}

A precise measurement of both total and differential cross sections would be definitely shed light on these issues,
particularly considering the different $t$-power behavior predicted by theories.
%
%
The state-of-the-art data are not accurate enough yet for a precise determination of the physics quantities \cite{Duran:2022xag},
mainly limited by the small rates for open and hidden charm photoproduction at threshold.
The Electron-Ion Collider in China(EicC) \cite{Anderle:2021wcy,Cao:2020NT,Cao:2020Sci}, as a next-generation dedicated experimental facility, provides a unique opportunity for exploring the the exclusive production of heavy quarkonium and exotic states in greater depth, which can unveil essential information about nucleon structure and the dynamics of quark-gluon interactions.
%
Several software packages, including SARTRE \cite{Toll:2012mb}, lAger \cite{Joosten:lAger}, and eSTARlight \cite{Lomnitz:2018juf,Klein:2019avl}, are available for simulating vector meson production in electron-ion scattering.
The exclusive charmonium production at these facilities has never been carefully investigated,
except for the $\Upsilon$ production under a preliminary design of detector at EIC-US \cite{Gryniuk:2020mlh}.
In this study, we employ eSTARlight to generate exclusive $J/\psi$ events, as depicted in Fig. \ref{fig:feynman}, under the EicC kinematic coverage.
This is achieved through the optimization of the input of photo-proton cross sections and the rectification of momentum reconstruction near the threshold.

We organize the rest of the paper as follows.
We describes how to generate the Monte Carlo events by considering the tracking efficiencies from a fast simulation of detector baseline design in Sec. \ref{sec:pseudodata}.
We explore the impact of those pseudo-data on some physics topics including GFFs and exotic states in Sec. \ref{sec:impact}.
In Sec. \ref{sec:sum} we briefly summarize our results.

\section{Pseudo-data Generation} \label{sec:pseudodata}

\subsection{Simulation Setup} \label{subsec:detector}

\begin{figure}
    \centering
    \includegraphics[scale=0.33]{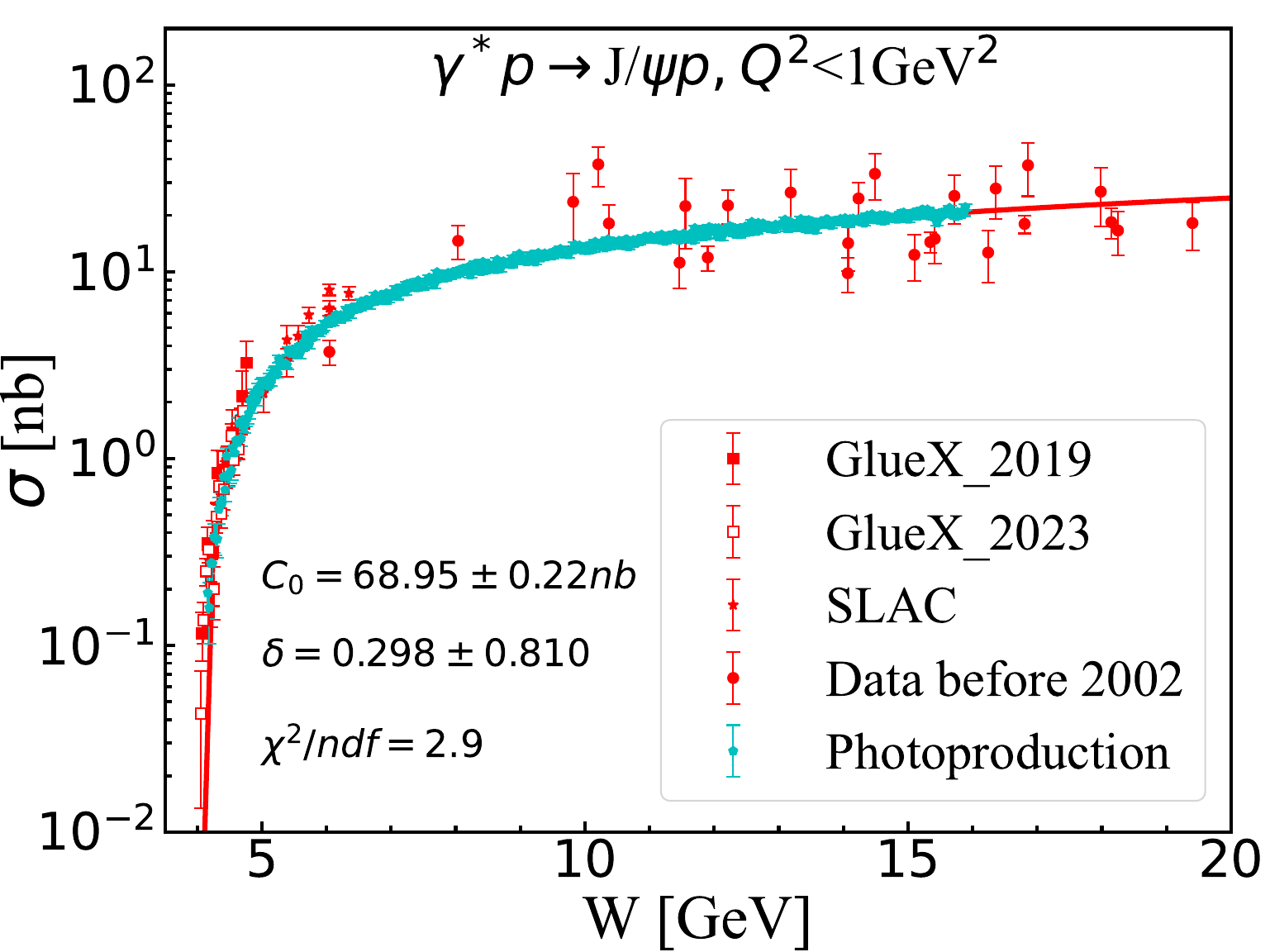}
    \caption{The total cross section of the $J/\psi$ exclusive photoproduction as a function of $W$ under $Q^2 < $ 1.0 {GeV}$^2$. The red solid line represents the parametrization in Eq. (\ref{eq:xsection}). The cyan points with small error bars represent the projection of $J/\psi$ photoproduction cross section within $W$ bins at the EicC. The experimental data are from Refs.\cite{E687:1993hlm,Amarian:1999pi,GlueX:2019mkq,GlueX:2023pev,EuropeanMuon:1979nky,Clark:1980ed,Clark:1979xs,Denby:1983az,Binkley:1981kv,Nash:1976ic}}.
    \label{fig:W_sigma}
\end{figure}

The eSTARlight utilizes a parameterization of the photoproduction cross-section of $\gamma p\to J/\psi p$ as an input of event generation.
The electroproduction cross sections in electron-proton scattering can be calculated with the help of photon flux of strong $Q^2$ dependence under equivalent photon approximation \cite{Lomnitz:2018juf,Klein:2019avl}.
A form factor is used to take account of the additional moderate $Q^2$ dependence of the interaction vertex.
The common longitudinal-to-transverse cross-section ratio in the literature~\cite{Martynov:2002ez,Martynov:2001tn} is used.

\begin{figure}
    \centering
    \includegraphics[scale=0.33]{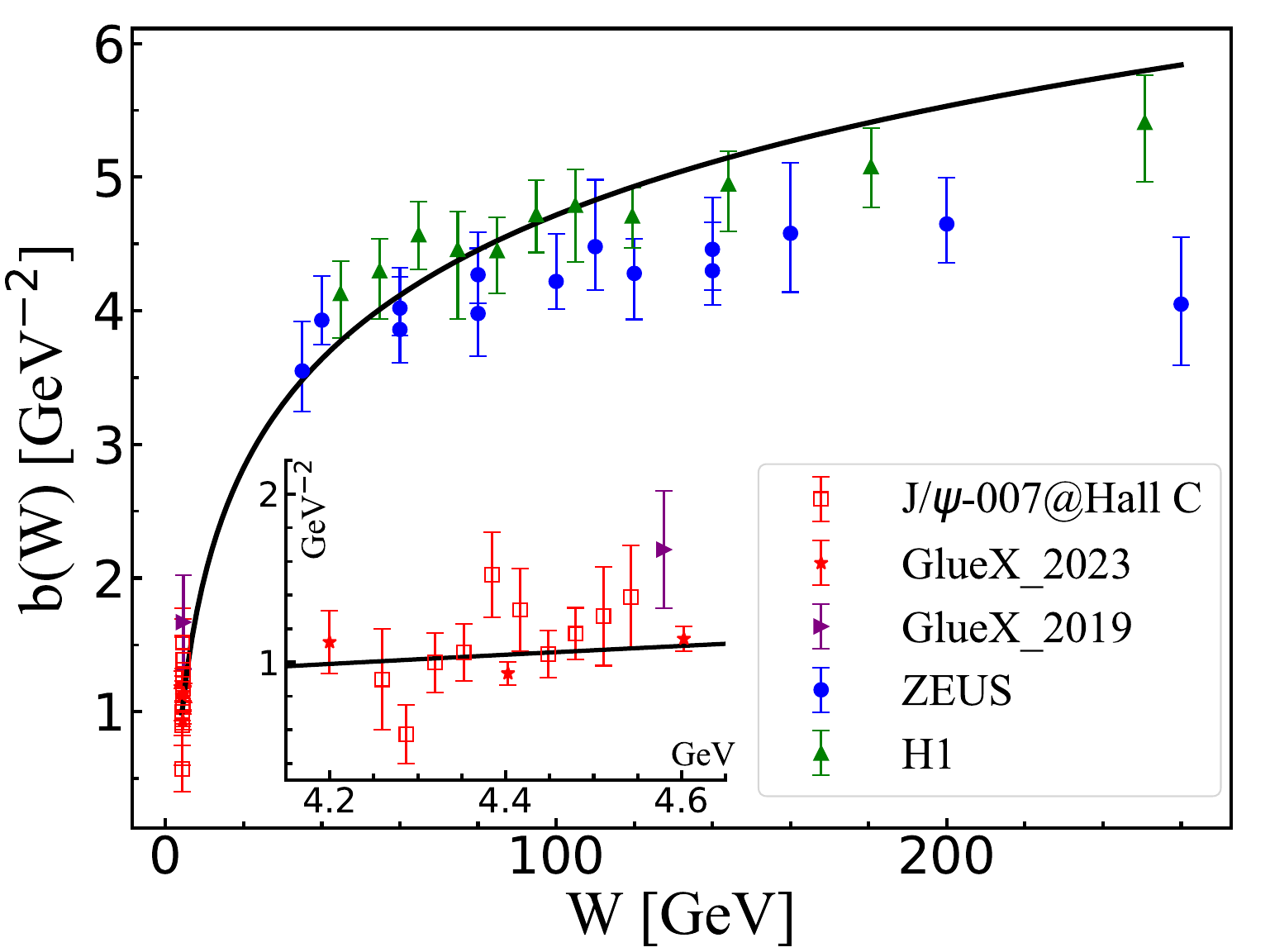}
    \caption{The slope parameter $b$ in Eq. (\ref{eq:slope}) as a function of $W$ in comparison of available data. Note that only the data below $W =$ 105 GeV is included in the fit. The inset is an enlarged of near-threshold region.}
    \label{fig:W_slope}
\end{figure}

Previously, the fitting of the $\gamma p\to J/\psi p$ was confined to high-energy data.
However, for our objectives, we have incorporated low-energy regime data using the following expression \cite{Cao:2018hvy}:
\be \label{eq:xsection}
\sigma(W)=C_0 \lf(1-\frac{(M_{p}+M_{\psi})^2}{W^2} \rg)^{1.5} \lf(\frac{W^2}{100^2 \text{ GeV}^2} \rg)^\delta
\ee
with proton mass $M_{p}$, $J/\psi$ mass $M_{\psi}$, and the invariant mass of the photon-
proton system $W$.
The parameters are determined by a fit to experimental data in the range of $4.05<W<100$ GeV, resulting into $C_0=68.95 \pm 0.22$ nb and $\delta=0.298 \pm 0.810$.
The fitting results are shown by the red curve in Fig. \ref{fig:W_sigma}.
The differential cross-sections are approximated by the exponential function $e^{-bt}$ (see Eq.~\ref{eq:exp}).
The slope parameter $b$ increases logarithmically with $W$ as inspired by the Regge phenomenology.
We adopt a prescription from a soft dipole Pomeron model with a double Regge pole in order to describe correctly the photo-Pomeron interaction in a wide energy range \cite{Martynov:2002ez,Martynov:2001tn}.
Within this framework, we add up one more Regge trajectory to incorporate the near-threshold behavior in addition to a hard Pomeron contribution:
\be \label{eq:slope}
b(W) = b_0 + 4 \alpha_0 \ln\frac{W}{4 \textrm{ GeV}} + 4 \alpha_1 \ln\frac{W}{90 \textrm{ GeV}}
\ee
with $\alpha_1 = 0.115 \textrm{GeV}^{-2}$ being determined from the fit to data at high energies \cite{ZEUS:2002wfj}.
To match the low energies data of GlueX \cite{GlueX:2019mkq,GlueX:2023pev} and $J/\psi$-007 \cite{Duran:2022xag} at JLab, we have identified $b_0 = 2.365 \pm 0.039 \textrm{GeV}^{-2}$, and $\alpha_0 = 0.178 \pm 0.008 \textrm{GeV}^{-2}$, a reasonable value in line with perturbative QCD expectation \cite{Frankfurt:2000ez}.
The fitting to the data of the energy-dependent slope $b$ up to $W =$ 105 GeV are shown in Fig.~\ref{fig:W_slope} with $\chi^2/\text{ndf} = 1.2$, which is sufficiently good for simulating case within the energy range of EicC.
The entire data sets of $t$-dependent cross-sections are present in the figure of Appendix \ref{apx:basics}.
The generated pseudodata with projected errors, that is driven by the inputting parametrization herein, will be compared to theoretical calculations whenever available in following sections for the purpose of scrutinizing whether the precision of future data could disentangle different models.

Another improvement of the event generator is on the near-threshold kinematics.
The approximate formula of a minimum of $|t|$ used in the original eSTARlight code is applicable only at high energies and becomes worse at low energies.
Thereby the momentum of generated events at low energies shall be upgraded to fulfill strictly the kinematic bound~\cite{Martynov:2001tn}:
\bea
t_0(y,Q^2) \geq &t& \geq t_1(y,Q^2) \nonumber
\eea
with the definition of
\bea
&& t_{0,1} = \nn \\ &&
\frac{(Q^2+M_p^2)(M_{\psi}^2-M_p^2)\pm \lambda(W^2, M_p^2, Q^2) \lambda(W^2, M_p^2, M_{\psi}^2)}{2 W^2} \nn \\ && - \frac{1}{2}(W^2+Q^2-M_{\psi}^2-2 M_p^2), \nonumber
\eea
where $Q^2$ is the virtuality of photon, and $\pm$ correspond to $t_{0}$ and $t_{1}$, respectively.
The K\"{a}ll\'{e}n triangle function is $\lambda^2(x,y,z) = x^2 + y^2 + z^2 -2xy - 2yz -2zx$.
%
Our construction procedure through energy-momentum conservation is illustrated as follows.
The momentum of the outgoing electron is determined by the sampled $Q^2$ and $W$ distributions since those of the initial beam particles are fixed by the facility design of EicC.
%
%
The momentum of the virtual photon is derived from the difference between the initial and final electron momentum, denoted as $q = p - p^\prime$.
%
%
Then the momenta of all particles in the final states are generated in the c.m. frame and transferred to the laboratory frame.
%
%
In these steps, there is no need for approximation as the conservation of momentum is maintained at each interaction vertex.
In the same spirit, an alternative prescription is employed by establishing the exact minimum $|t|$ \cite{Li:2022kwn}.


Following an update to the cross-section parameterization and a correction in kinematic reconstruction, the eSTARlight package is now capable of generating $J/\psi$ exclusive events from near-threshold regions to high energies.
This updated package is utilized to determine the reconstruction resolution and efficiency in subsequent detector performance studies.

\begin{figure*}[htbp]
   \centering
   {\includegraphics[width=1.0\linewidth]{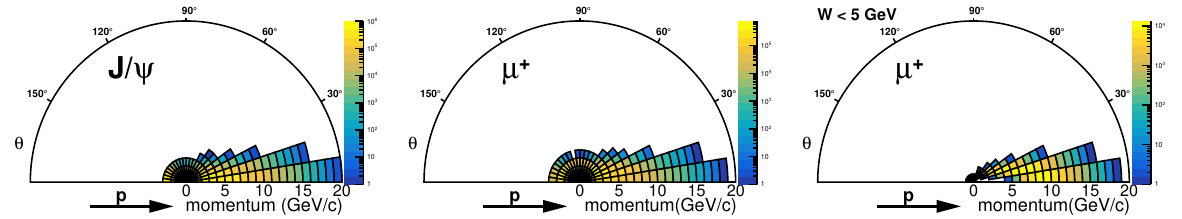}}
   \caption{Momentum (radial) and polar angle (polar) distributions for $J/\psi$ and its daughter lepton in the laboratory frame in the range of $Q^2 <$ 1.0 GeV$^2$. The third panel is restricted to the range of $W < 5$ GeV. The distributions of antimuon are nearly the same and not shown herein.}
   \label{fig:kinematic_map_mumu_Q2<1_all_combine}
\end{figure*}

The initial conceptual design for the EicC detector is outlined in the white paper~\cite{Anderle:2021wcy}.
As a general-purpose detector designed for asymmetric collisions between electron beams and proton/ion beams, the design incorporates more materials on the hadron/ion-going side.
The detector is composed, from the inside out, of the vertex/tracking detector, the particle identification system, and the calorimeter system, among others \cite{He:2023svm,He:2024ime}.
The crossing angle between the colliding beams are 50 milli radian by the design of the interaction point.

The current design of the tracking system employs hybrid models.
For middle rapidity ($|\eta|<1.1$), there are five silicon layers and four Micro-Pattern Gaseous Detector (MPGD) layers, extending radially from 3.3 cm to 77.5 cm.
For $|\eta|>1.1$, the tracking system comprises silicon disks succeeding by large-area Micromegas in the forward (proton/nucleus going) direction and solely silicon disks in the backward (electron-going) direction.
A comprehensive GEANT4 simulation has been conducted for this tracking configuration, and the resolutions for primary vertex position, the distance between tracks to the interaction point, and track momentum, as well as the tracking detector efficiency as a function of track transverse momentum $p_T$ and pseudorapidity $\eta$, are elaborated in Ref.~\cite{Anderle:2023uvi}.
A fast simulation framework has also been developed to simulate the detector responses derived from the GEANT4-based simulation.
In this study, we adhere to the same fast simulation procedure described in Ref.~\cite{Anderle:2023uvi}.

The final $J/\psi$ is reconstructed through its dilepton decay, including both the $\mu^{+}\mu^{-}$ and $e^{+}e^{-}$ channels.
Momentum (radial) versus polar angle (polar) distributions for $J/\psi$ and its daughter dilepton in the laboratory frame are shown in Fig. \ref{fig:kinematic_map_mumu_Q2<1_all_combine}.
The distributions of electron and positron are very similar to those of dimuon with only minor differences.
%
%
The $J/\psi$ particle is predominantly produced in the direction of the proton beam, with a significant portion emerging at a very forward angle.
According to our Monto Carlo simulation, the momentum resolution is excellent (below 1\%) in a wide range of electron momentum (up to 15 GeV when $|\eta|<2.5$), resulting into a fine mass resolution of the reconstructed $J/\psi$. The good mass resolution enables an effectively discriminates the genuine $J/\psi$ signals from the background events.
In the following analysis, only the $\mu^{+}\mu^{-}$ channel is considered for statistical uncertainties, taking into account the nearly identical efficiency for electrons and muons.
By further including and $e^{+}e^{-}$ channels, the statistical errors in the subsequent simulations is reduced by a factor of $\sqrt{2}$.
However, one should keep in mind that the influence of bremsstrahlung when interacting with the material electron traverses has not yet been considered \cite{ZEUS:2002wfj}.

\begin{figure}
    \centering
    \includegraphics[scale=0.33]{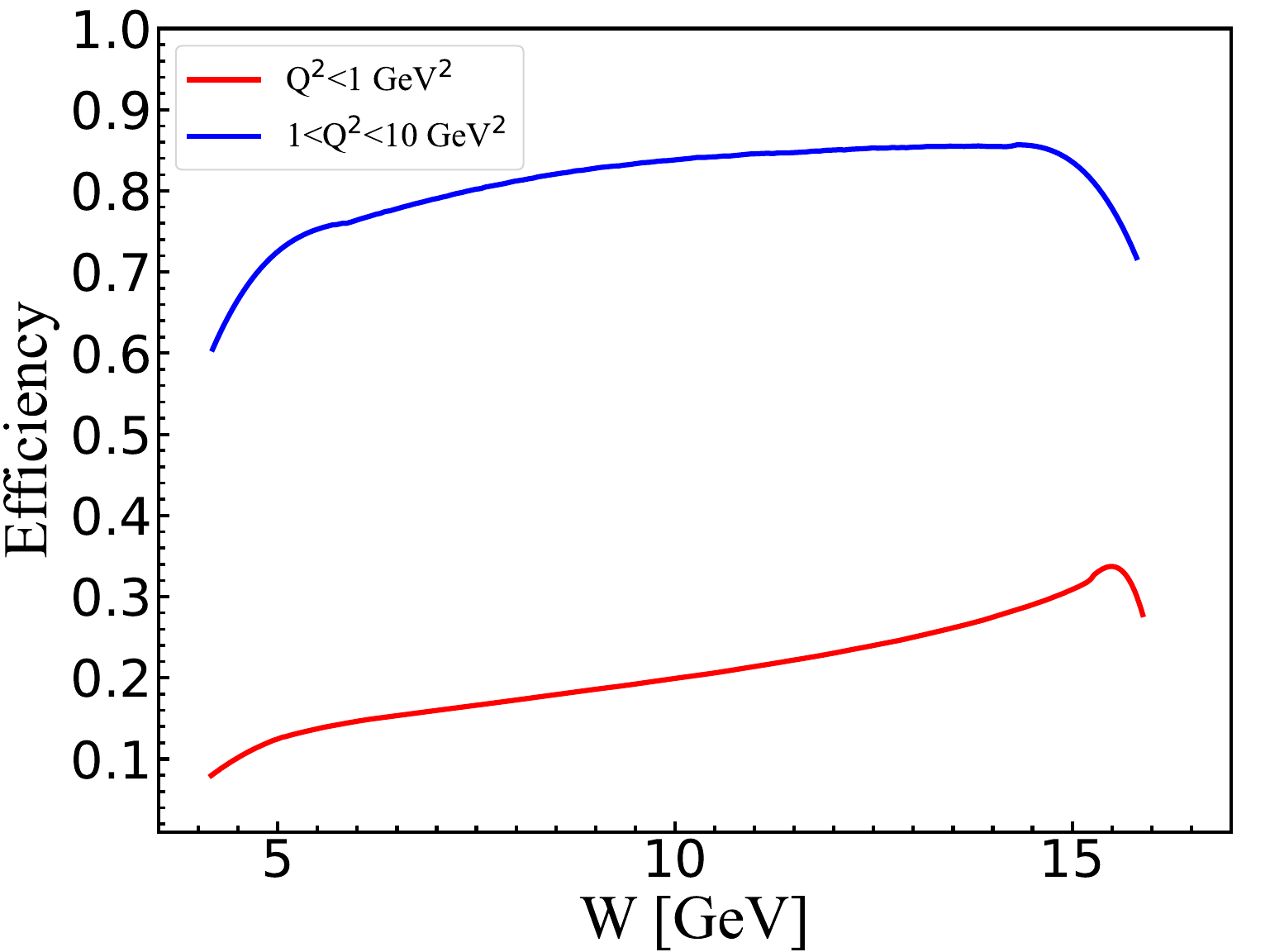}
    \caption{Overall detection efficiency of $ep \rightarrow ep J/\psi \rightarrow ep \mu^{+}\mu^{-}$ as a function of $W$. Efficiency here is the ratio of the number of reconstructed $J/\psi$ events by detector to generated ones by eSTARlight.
    The $Q^2$ bins are separated as $Q^2 <$ 1.0 GeV$^2$ (red curve) and 1.0 $< Q^2 < $ 10.0 GeV$^2$ (blue curve).}
    \label{fig:W_efficiency}
\end{figure}

The resolution of $W$ is examined by reconstructing it from the momentum of the final dilepton and recoil proton, considering a momentum resolution of 0.1\%.
As a result, the resolution of $W$ is 5.5 MeV above $W = 10.0$ GeV, 14 MeV in the interval of 5.0 GeV $< W <10.0$ GeV, and 27 MeV below $W = 5$ GeV.
The resolution of $W$ deteriorates at the threshold region because the decaying dilepton is off central rapidity (see the third panel in Fig. \ref{fig:kinematic_map_mumu_Q2<1_all_combine}) though the recoil proton is of good momentum resolution.
%
%
%
Therefore, the need for a far-forward detection system in the direction of the proton beam is identified, which will require further optimization in the future.
The minimum $W$ is limited to 4.16 GeV by the statistics after considering detection acceptance and efficiency of the detector.
The overall reconstruction efficiency as a function of $W$ is shown in Fig. \ref{fig:W_efficiency}.
It is noticed that for quasi-real photon ($Q^2 <$ 1.0 GeV$^2$) the scattered electrons are boosted to negative pseudo-rapidity.
%
%
The efficiency near the threshold, as shown by the red curve in Fig. \ref{fig:W_efficiency}, is still limited,
indicating a future installation of the far-forward detector in the direction of the electron beam.
For deep inelastic physics requiring $Q^2$ larger than 1.0 GeV$^2$, a detector coverage of $\eta >$ -3 is sufficient for the scattered electron.
%
%
The overall efficiency within the region of 1.0 $< Q^2 < 10$ GeV$^2$, as shown by the blue curve in Fig. \ref{fig:W_efficiency}, is significantly higher than that of $Q^2 <$ 1.0 GeV$^2$, as anticipated.

\subsection{Projection pseudodata at the EicC} \label{subsec:bins}

%
The nominal configuration for the EicC proposal involves a collider with a 3.5 GeV electron beam and a 20 GeV proton beam.
This corresponds to a center of mass energy ($\sqrt{s}$) of $16.76$ GeV.
We assume an integrated luminosity of 50 fb$^{-1}$, corresponding to 289 days running under the designed luminosity $2 \times 10^{33} \text{cm}^{-2}\textrm{s}^{-1}$.
%
%
The quasi-real cross-section of $ep \rightarrow epJ/\psi \to ep \mu^+ \mu^-$ calculated by eSTARlight generator is 42.39 pb and the raw events are 2119500.
The same width 25 MeV of $W$ bin is used in our analysis, resulting into a total of 480 bins for $Q^2 < $ 1.0 {GeV}$^2$.
%
%
The cyan points in Fig. \ref{fig:W_sigma} represent the pseudo-data generated for the cross-section within each $W$ bin, filtered by the detector designed for EicC.
For the high-$W$ region of the quasi-real domain, the statistical errors are very small and the systematic ones will dominate the true measurements in the future.
The statistic errors increase rapidly to be around 10\% below $W = 4.48$ GeV because of the phase space suppression and the detector limitation.

\begin{figure}
    \centering
    \includegraphics[scale=0.33]{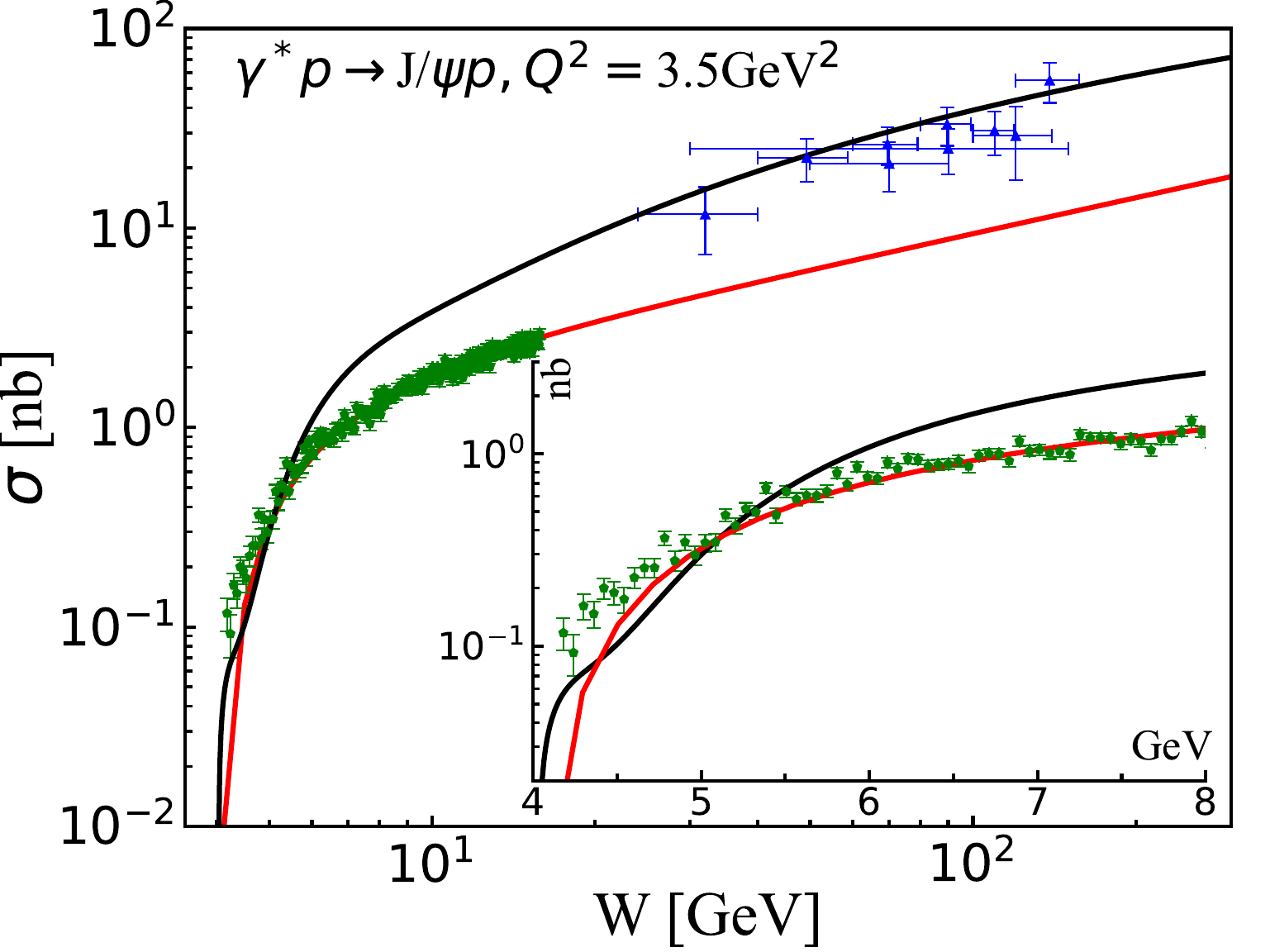}
    \caption{The total cross section of the $J/\psi$ exclusive photoproduction as a function of $W$ in the weighted average $Q^2 = 3.5 \text{GeV}^2$ between the interval 1.0 $< Q^2 < $ 10.0 GeV$^2$. The green points are the projection of $J/\psi$ photoproduction cross section in each $W$ bin at the EicC. The blue triangles are the ZEUS data \cite{ZEUS:2002wfj}. The black solid line is the result of a soft dipole Pomeron model fitting to worldwide data \cite{Martynov:2002ez}. The red solid line represents the input model curve. The inset is an enlarge of near threshold region.}
    \label{fig:W_sigma_Q2}
\end{figure}

In Fig. \ref{fig:W_sigma_Q2}, the generated data of cross-sections with respect to $W$ at 1.0 $ < Q^2 <$ 10.0 GeV$^2$ are shown together with the ZEUS data \cite{ZEUS:2002wfj}.
The total cross-section of $ep \rightarrow epJ/\psi$ is 3.142 pb and the raw event number is 157100.
%
%
Considering the statistics and the detector resolution, the entire $W$ coverage is divided into 195 bins, each with a width of 60 MeV.
The black solid line represents the fit of a soft dipole Pomeron model to the world data at that time \cite{Martynov:2002ez}, while the red solid line represents the input model curve.
%
%
The relative statistic errors are well below 10\% above $W = 4.60$ GeV  and increase rapidly to be around 15\% when approaching closer to the threshold.

\begin{figure*}[htbp]
   \centering
   {\includegraphics[width=0.92\linewidth]{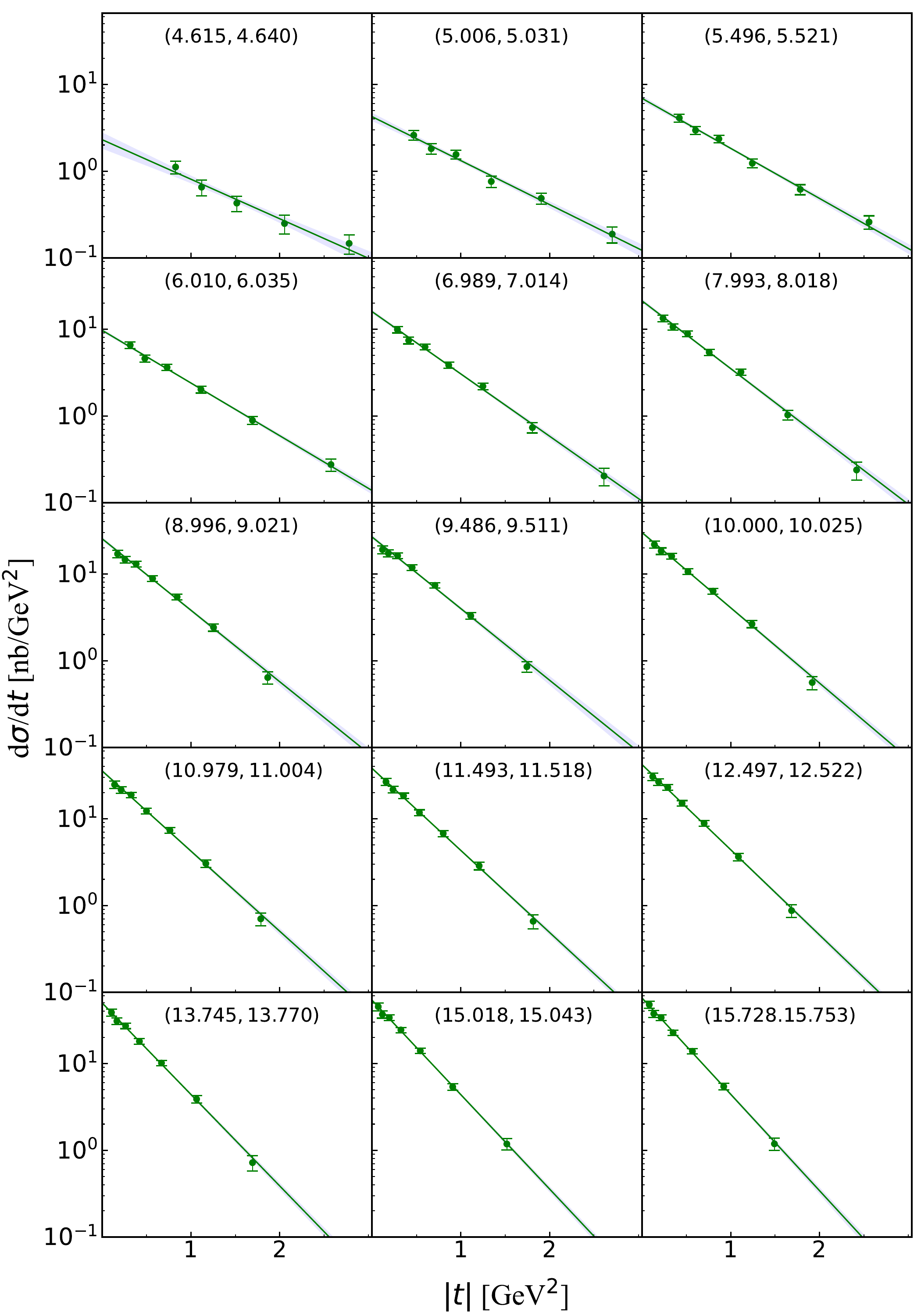}}
   \caption{The $t$-dependence of the $\gamma^* p \rightarrow J/\psi p$ differential cross section for $Q^2 <$ 1.0 GeV$^2$ in selected $W$ bins as indicated inside the parentheses on the subplot.
   The bands represent the uncertainty propagated in a fit of an exponential $t$-dependence. }
   \label{figs:EicC_dsigmadt_combined_plots}
\end{figure*}

\begin{figure}[htbp]
   \centering
   {\includegraphics[width=0.95\linewidth]{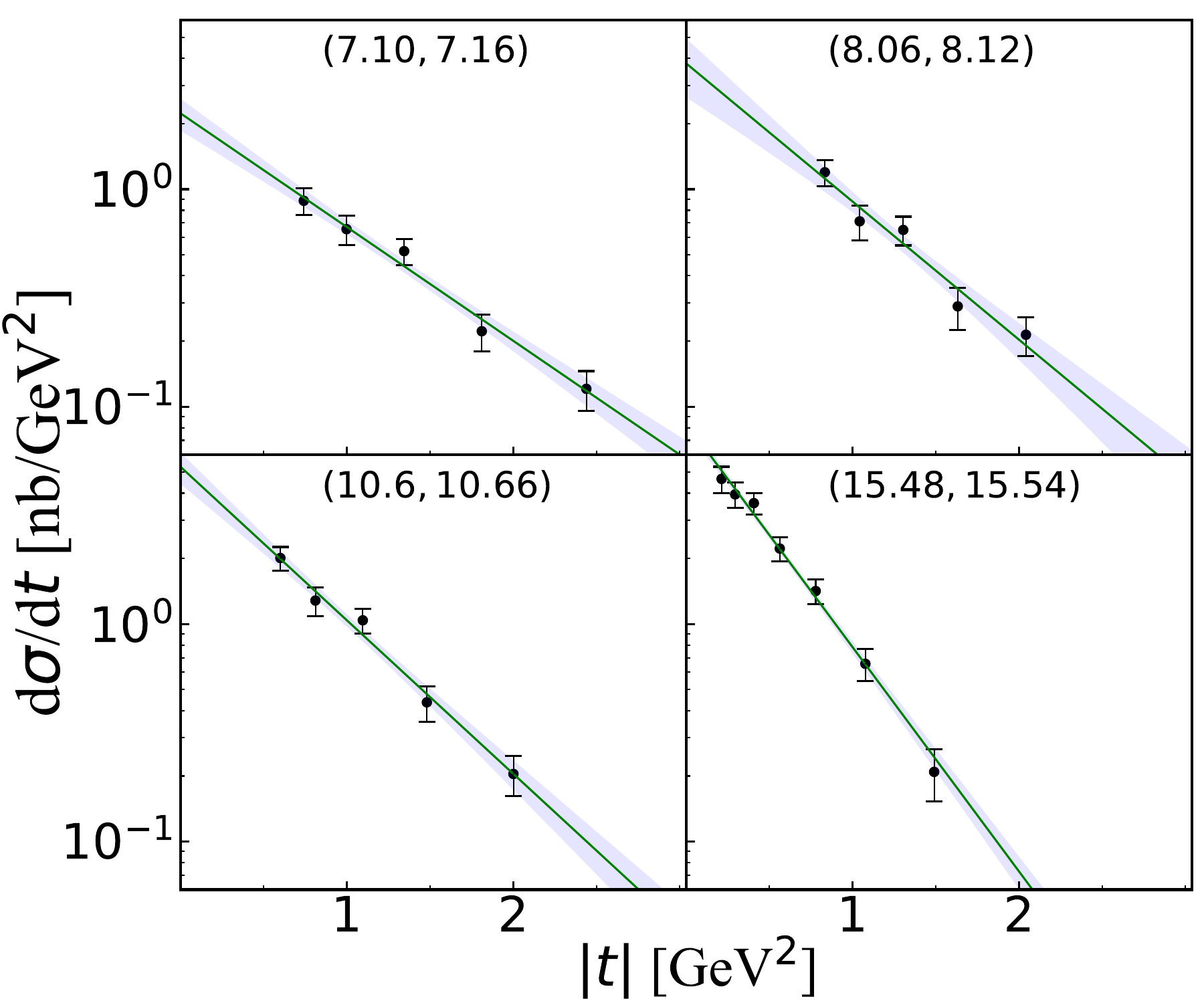}}
   \caption{The $t$-dependence of the $\gamma^* p \rightarrow J/\psi p$ differential cross section for 1.0 $< Q^2 < $ 10.0 GeV$^2$ in selected $W$ bins as indicated inside the parentheses on the subplot. The bands represent the uncertainty propagated in a fit of an exponential $t$-dependence.}
   \label{figs:EicC_dsigmadt_Q2_3.5}
\end{figure}

The Fig.~\ref{figs:EicC_dsigmadt_combined_plots} and Fig.~\ref{figs:EicC_dsigmadt_Q2_3.5} show the simulated results for the $t$-dependent cross sections for selected $W$ bins within the range of $Q^2 < 1.0$ {GeV}$^2$ and 1.0 $ < Q^2 <$ 10.0 GeV$^2$, respectively.
It is seen that the maximum $t$ covered by the detector is around 2.5 GeV$^2$.
%
%
For the region close to the threshold, we need to combine the bins to increase the statistics.
This results in the minimum $W$ value for the measurement of differential cross-sections rising to 4.35 GeV.
%

The line shape of cross section of $J/\psi$ production is expected to depend on the specific production mechanism, which would contribute to different energies and/or kinematic regions.
Whether the precision of future data could disentangle them is our main concern here, and next section will discuss some of the selected topics by comparing the projected uncertainties to the model calculations.

\section{Insight into the relevant physics} \label{sec:impact}

\begin{figure}
    \centering
    \includegraphics[scale=0.33]{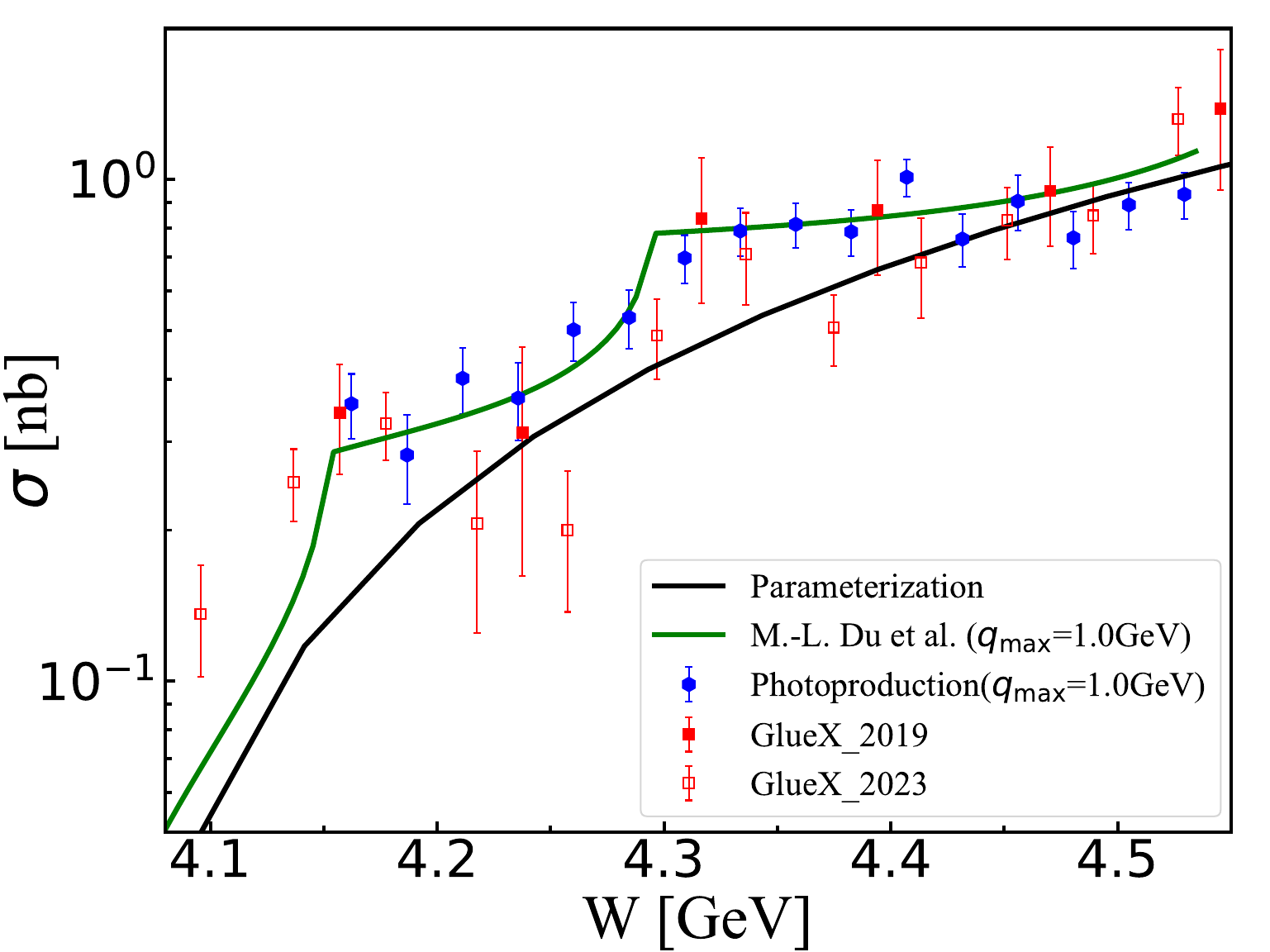}
    \caption{The near-threshold pseudodata (blue dots) at EicC generated based on the theoretical calculation (green solid curve) with a cut-off value $q_{\textrm{max}} = 1.0$ GeV in the box diagrams \cite{Du:2020bqj}.
    The red data points are from GlueX \cite{GlueX:2019mkq,GlueX:2023pev}.
    The black curve is the parameterization in Eq.~(\ref{eq:xsection}).
}
    \label{fig:W_nearthreshold_qmax_1}
\end{figure}

\begin{figure}
    \centering
    \includegraphics[scale=0.34]{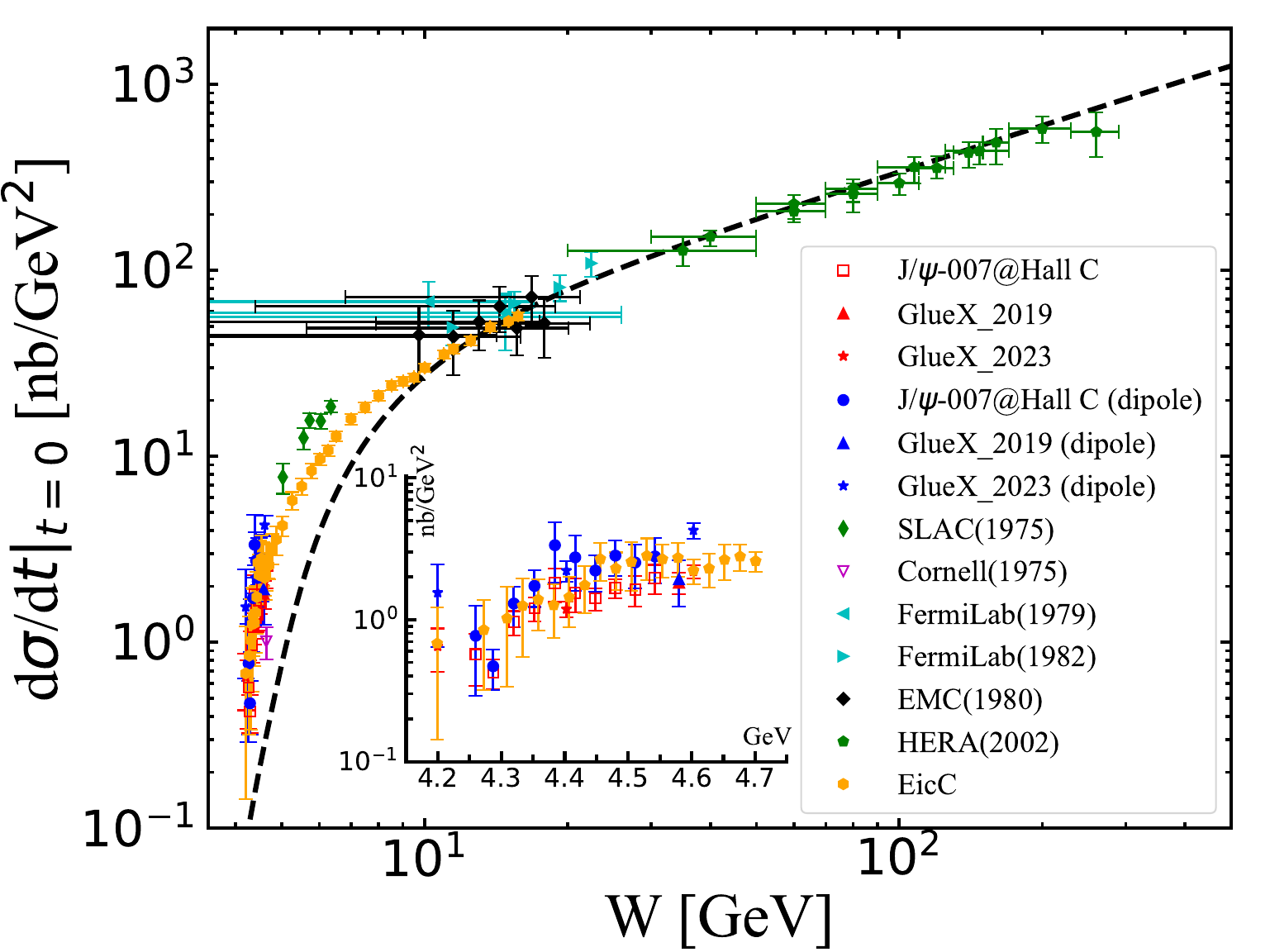}
    \caption{The differential cross section of $\gamma^* p \to J/\psi p$ in the forward direction ($t$ = 0) as a function of $W$ by the extrapolated method.
    The orange data points are obtained from the pseudodata at EicC.
    The blue data points are extrapolated of GlueX data by dipole formula, and the other points are those by exponential function.
    The experimental data are from \cite{Duran:2022xag,GlueX:2019mkq,GlueX:2023pev,Camerini:1975cy,Gittelman:1975ix,EuropeanMuon:1979nky,ZEUS:2002wfj,Binkley:1981kv}.
    The black dashed line is the contribution of imaginary part of amplitude through a once-subtracted dispersion relation \cite{Gryniuk:2016mpk}.
    The inset is an enlarge of near threshold region. }
    \label{fig:W_dsigmadt_0}
\end{figure}

The precise measurement of $J/\psi$ exclusive photoproductuon at low energies is of fundamental interest in the underlying dynamics.
The well-established pentaquarks below 4.5 GeV are narrower than our $W$ resolution as seen by the LHCb \cite{Aaij:2015tga,Aaij:2019vzc}.
Their possible signal featured by Breit-Wigner line shapes of $J/\psi p$ would be broadened or even disappear by convolutional folding of $W$ resolution.
The excitation of pentaquark states is expected to be of a mass bigger than 5.0 GeV, which could be searched for because of the excellent $W$ resolution at higher energies.
If pentaquarks indeed have sizable branching ratios in the $J/\psi p$ channel, as predicted by theories \cite{Wu:2010jy,Huang:2018wed}, both differential and total cross section data could be used to explore the small photocoupling in their radiative decay, or even to quantitatively justify the VMD assumption \cite{Cao:2023rhu,JointPhysicsAnalysisCenter:2023qgg}.
The GlueX data \cite{GlueX:2023pev} have already witnessed faint signs of trend beyond the exponential behavior in the range of $-t > 3$ GeV$^2$ (see lower panel in Fig. \ref{fig:overall} of Appendix \ref{apx:basics}), unfortunately not covered by the present detector design at EicC.
The pasudodata of 1.0 $< Q^2 < $ 10.0 GeV$^2$ in Fig. \ref{figs:EicC_dsigmadt_Q2_3.5} shows the possibility of further exploring these issues through the electromagnetic transition of the amplitudes.

Within perturbative QCD, the production amplitude near the threshold is factorized with regard to the gluonic generalized parton distributions and the quarkonium (herein $J/\psi$) distribution amplitude in the domain of large-$t$ and big skewness with a definition of \cite{Guo:2021ibg,Guo:2023pqw,Guo:2023qgu}:
\be
\xi = \frac{t - M_{\psi}^2}{2 M_p^2 + M_{\psi}^2 -t - 2 W^2}
\ee
The pseudodata of $t$-dependent cross sections at EicC are distributed in the range of $0.33 <\xi < 0.50$, among of which only 31 data points with moderate statistical errors satisfy the requirement $\xi > 0.4$ of pQCD expansion.
So a fair impact on extraction of GFFs is anticipated in the leading moment approximation \cite{Guo:2021ibg} or holographic QCD \cite{Mamo:2021krl}, however, of some model dependence \cite{Sun:2021pyw,Sun:2021gmi}.

Additionally, from another perspective, the near-threshold dynamical mechanism could be completely different from that based on gluon exchange, as suggested by M.-L. Du \textit{et al.} \cite{Du:2020bqj}.
The $J/\psi p$ final states are possibly produced by the rescattering through box diagrams via $\Lambda_c \overline{D}$ and $\Lambda_c \overline{D}^{\ast}$ open-charm intermediate states.
The cusp structures at the thresholds are proposed as a distinguishing feature so can be used as a means to confirm or deny this mechanism by precise measurement of total cross sections.
The blue dots in Fig.~\ref{fig:W_nearthreshold_qmax_1} present the simulated statistical uncertainties of the $J/\psi$ photoproduction cross section near the threshold at EicC.
The future data can be used to disentangle the second cusp stemming from the $\Lambda_c \overline{D}^{\ast}$ rescattering, therefore, scrutinizing different scenarios.
Whether the data, particularly the near-threshold differential cross sections, allow a distinction among the soft pomeron exchange, two-gluon and three-gluon exchange needs more theoretical input.

The $t$-dependent cross sections are a further tool to access the production
mechanisms \cite{JointPhysicsAnalysisCenter:2023qgg}, with a particular interest in the forward direction.
The minimum $|t|$ can reach very close to zero at high energies and
distance from zero if approaching closer to the production threshold.
Therefore the extraction of forward cross-sections at low energies relies on the extrapolation.
The Fig. \ref{fig:W_dsigmadt_0} illustrates the $W$-dependent differential cross sections of the $\gamma p \rightarrow J/\psi p$ at forward direction by exponential extrapolating to $t=0$.
Both pesudodata at EicC and available experimental data are shown together with the contribution of the imaginary part of amplitude through a once-subtracted dispersion relation (black dashed line) \cite{Gryniuk:2016mpk}.
The systematic errors of extrapolation have not been studied in detail. However, an extrapolation of GlueX data using the dipole formula (see Eq. (\ref{eq:dipole})) is shown for comparison. The errors in this case are larger than those of the exponential extrapolation.
More precise measurements of wider kinematic coverage will definitely decrease the uncertainties of extrapolation. 
It is clearly revealed that those data of moderate precision can be used to probe the real part of $J/\psi p$ production mechanism at close-to-threshold region, thus validate different theoretical models.
How to relate these data to underlying physics, e.g. the $J/\psi$-p scattering length \cite{Gryniuk:2016mpk}, calls for more theoretical works, of particular attention to the controversial issue on VMD assumption.

\section{Summary and conclusion} \label{sec:sum}

Physics opportunities with exclusive photo- and electro-production of heavy quarkonium covers multiple interesting scope.
The apparatus at JLab measures the near-threshold cross sections with a wide kinematic coverage compared to the past facilities.
Whether the future facilities can explore broader kinematic regions with high precision needs a careful scrutinizing. 
Our paper uncovers the feasibility of $J/\psi$ exclusive production by a state-of-the-art design of detector at EicC.
With the aim of more reliable sampling of the $ep \rightarrow epJ/\psi$ process, the model components are updated to generate the events within eSTARlight generator by considering the recent data set from JLab.
Besides, a correction of kinematic reconstruction near the threshold allows a more trustworthy simulation of the momenta of final particles.
The total and differential cross sections with statistical uncertainties of the $ep \to eJ/\psi p$ process are projected at the designed energy of EicC under the integrated luminosity of 50 fb$^{-1}$.
The precision of total cross sections in the quasi-real regime enables a sensitive reconciling of the rear-threshold mechanism with an emphasis on the cusp of $\Lambda_c \overline{D}^{\ast}$ threshold.
The $t$-dependent cross sections at big skewness can be used to constrain the GFFs, unveiling the mechanical properties of proton \cite{Kharzeev:2021qkd}.
At high energies, the remarkably precise measurements at EicC fill the $W$ gap between HERA and JLab data, e.g. 6.5 $ <W <$ 9.5 GeV, and the $t$-dependence in the range of 6.5 $ <W <$ 16.755 will be measured with narrow bin width and high precision.
Sufficient statistics and detector resolution allow for narrow $W$ binning while maintaining small statistical uncertainties. This therefore has the potential to discover the narrow signal of pentaquark excitation amidst the continuum of gluon exchange.
The accurate measurements of high $Q^2$ region will impose a powerful investigation of electromagnetic transition of amplitudes at the close-to-threshold region, and possibly the gluon part of the D-term GFFs \cite{Boussarie:2020vmu}.
In a word, our results of the sensitivity projection reveals the kinematic coverage and precision of charmonium production under a realistic design of the detector at future colliders.


The event rates across the entire kinematic range serve as vital benchmarks for design and optimization of the detector at the EicC, particularly for the data collection and reconstructing strategies of exclusive processes.
The performance of the proposed detection system for recoil protons, final $J/\psi$, and decaying leptons is excellent at high $W$ and $Q^2$. However, the efficiency and resolution of detectors pose challenges when approaching the near-threshold region and large-$|t|$ domain. The resolution and the minimum achievable value of $W$ are limited by the detection of events with large pseudorapidity. Therefore, there is still room for improvement of the far-forward detector by optimizing the reconstruction scheme and tracking strategies.
Other exclusive $J/\psi$ processes through the $t$-channel in electron-ion scattering, such as the $e^-p \rightarrow e^-p J/\psi \pi$ and $e^-p \rightarrow e^-p J/\psi \pi \pi$ for the study of exotic states $Z_c(3900)$ and $X(3872)$ respectively, can also be simulated by the eSTARlight package \cite{Xie:2020wfe,Xie:2020ckr}.
Together with other essential exclusive processes such as $\eta_c$-meson \cite{Jia:2022oyl,Benic:2023ybl}, $\Upsilon$-meson production \cite{Cao:2019gqo,Gryniuk:2020mlh}, and Deeply Virtual Compton Scattering \cite{Cao:2023wyz}, the precision studies of these exclusive processes are not only beneficial for the design of EicC, but also provide insights into the performance of the low energy domain of the EIC \cite{Burkert:2022hjz}.

\bigskip

\begin{acknowledgments}

Useful discussions with Hongxin Dong, Y Hatta, Xinbai Li, Peng Sun, Xin Wu, Feng Yuan, and Wangmei Zha are gratefully acknowledged.
Special thanks go to Feng-Kun Guo for help in generating Fig. \ref{fig:W_nearthreshold_qmax_1} and Weizhi Xiong for help in design of the detector.
This work is supported by the National Natural Science Foundation of China (Grants Nos. 12075289, U2032109, 11905092, 12105132 and 11705078); The Strategic Priority Research Program of Chinese Academy of Sciences (Grant NO. XDB34030301); Opening Foundation of Songshan Lake Materials Laboratory, Grants No.2021SLABFK04.

\textbf{Data Availability Statement}
This manuscript has no associated data or the data will not be deposited. [Authors’ comment:The pseudo-datasets generated during and/or analysed during the current study are available from the corresponding author on reasonable request.]


\end{acknowledgments}

\bigskip

\appendix

\section{The fit to the data at low energies } \label{apx:basics}

\begin{figure*}[htbp]
   \centering
   {\includegraphics[width=0.83\linewidth]{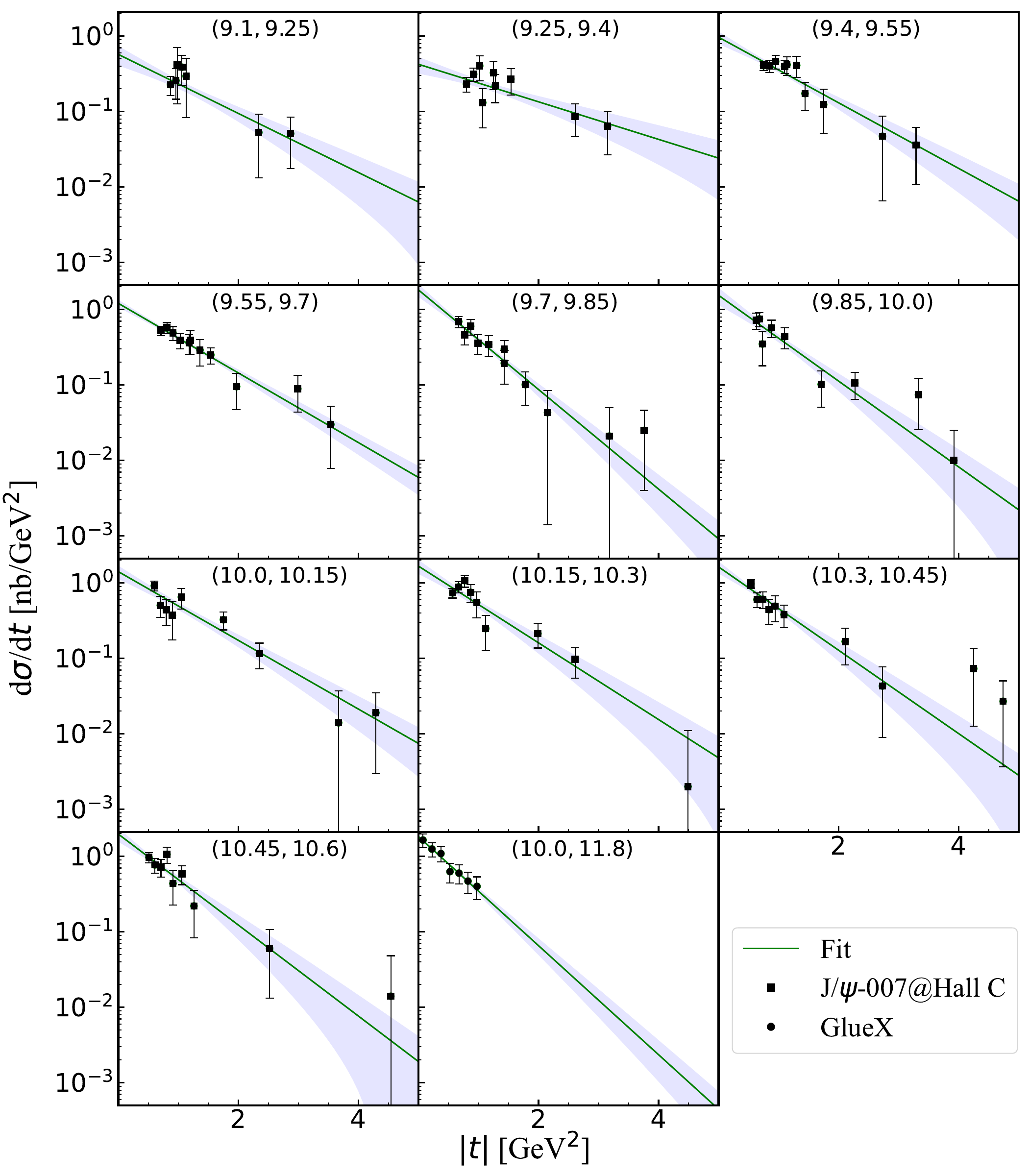}}
   {\includegraphics[width=0.84\linewidth]{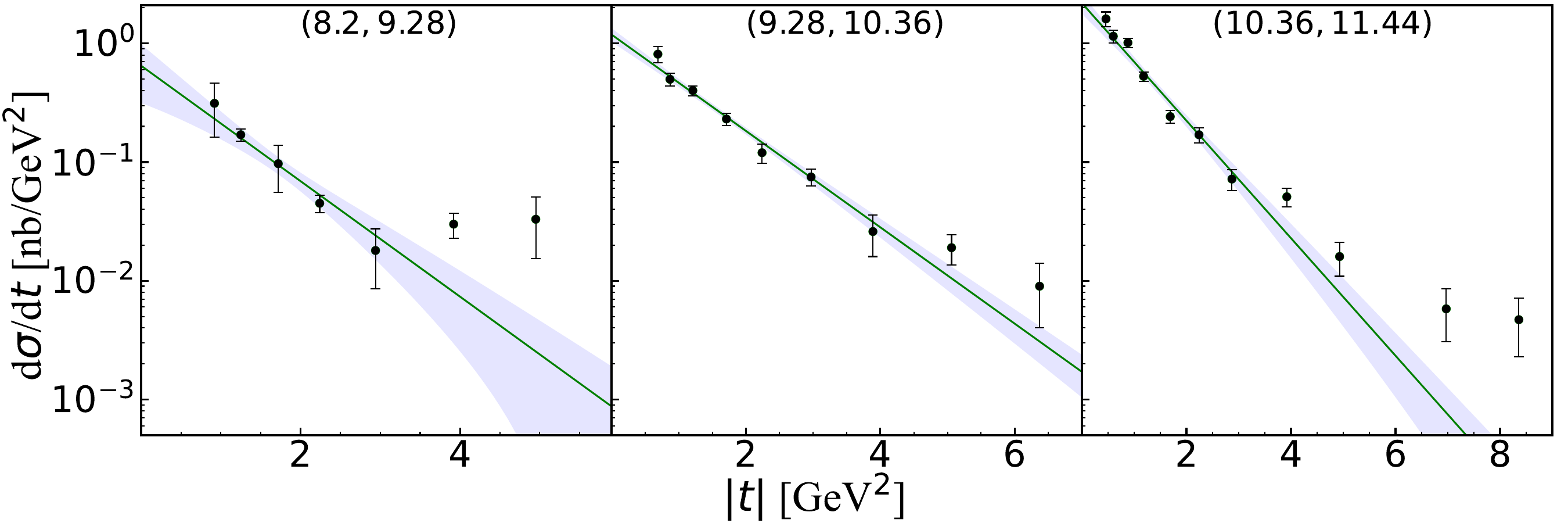}}
   \caption{The differential cross sections of the $J/\psi$ photoproduction as a function of $-t$. The numerical values inside the parentheses in each subplots indicate the range of photon energy in the rest frame of the initial proton $E_{\gamma} = (W^2 - M_{p}^2)/{2 W}$ in the unit of GeV.
   The experimental data are from GlueX \cite{GlueX:2019mkq,GlueX:2023pev} and $J/\psi$-007 \cite{Duran:2022xag} at JLab. }
   \label{fig:overall}
\end{figure*}

In Fig.~\ref{fig:overall} we show the entire data sets of $t$-dependent cross sections together with our fit  by use of
\be \label{eq:exp}
\frac{d\sigma}{dt}(\gamma p \rightarrow J/\psi p) \propto  e^{-bt}
\ee
The extracted slope $b$ in each subplot are shown in Fig. \ref{fig:W_slope}.
The forward differential cross sections of $\gamma^* p \to J/\psi p$ are shown in Fig. \ref{fig:W_dsigmadt_0} by extrapolating to the $t$ = 0.
Alternatively a  dipole formula is used for GlueX data in Fig. \ref{fig:W_dsigmadt_0}:
\be \label{eq:dipole}
\frac{d\sigma}{dt}(\gamma p \rightarrow J/\psi p) \propto  \frac{1}{\lf( 1 - \frac{t}{\Lambda^2} \rg)^2}
\ee

\bibliography{HFatEicC_ref}

\end{document}